\documentclass[twocolumn,showpacs,preprintnumbers,amsmath,amssymb]{revtex4}
\usepackage{graphicx}
\usepackage{dcolumn}
\usepackage{bm}

\begin{document}

\preprint{APS/123-QED}

\title{Quantum asymmetric cryptography with symmetric keys}

\author{Fei Gao$^{1,2}$, \quad Qiao-Yan Wen$^{1}$, \quad Su-Juan Qin$^{1}$, and Fu-Chen Zhu$^{3}$\\
        (1. State Key Laboratory of Networking and Switching Technology, Beijing University of Posts and \\ Telecommunications, Beijing, 100876, China) \\
        (2. State Key Laboratory of Integrated Services Network, Xidian University, Xi'an, 710071, China)\\
        (3. National Laboratory for Modern Communications, P.O.Box 810, Chengdu, 610041, China)}

\date{\today}

\begin{abstract}
Based on quantum encryption, we present a new idea for quantum
public-key cryptography (QPKC) and construct a whole theoretical
framework of a QPKC system. We show that the quantum-mechanical
nature renders it feasible and reasonable to use symmetric keys in
such a scheme, which is quite different from that in conventional
public-key cryptography. The security of our scheme is analyzed
and some features are discussed. Furthermore, the state-estimation
attack to a prior QPKC scheme is demonstrated.
\end{abstract}

\pacs{03.67.Dd, 03.67.Hk, 03.65.Ud}

\maketitle

\section{Introduction}
In the 1970s, the concept of public-key cryptography (PKC), also
called asymmetric cryptography, was proposed \cite{DH76,RSA78}. It
represented the radical revision of cryptographic thinking and
transformed the world of information security. Before the
appearance of PKC, the tool for keeping the secrecy of
communications was symmetric cryptography, where two parties
involved in the communication must previously share a sequence of
secret bits (i.e., the key) to encrypt and decrypt the message. In
this condition how to securely distribute such a key between the
users becomes an intractable problem. On the contrary, in PKC
there are two \emph{different} keys $e$ and $d$ (this is the
reason why it is also known as \emph{asymmetric} cryptography),
called the public key and the private key respectively. Just as
their names imply, $e$ would be published and anyone can access it
freely, whereas $d$ is only known to its owner. As described by
Rivest, Shamir, and Adleman when they presented the famous RSA
scheme \cite{RSA78}, a PKC system generally satisfies the
following four conditions: (C1) A message encrypted with $e$ can
be correctly decrypted with $d$; (C2) Both the encryption and the
decryption are easy to compute; (C3) It is difficult to compute
$d$ from the public $e$; (C4) A message encrypted with $d$ can
also be correctly decrypted with $e$. Armed with these properties,
PKC can be conveniently utilized by users, who do not need
previously share a secret key anymore. Therefore, PKC can resolve
the difficulty of key distribution in symmetric cryptography, and
then the latter can be used to encrypt the messages. This kind of
hybrid cryptosystem is generally used in our practical
implementations. Furthermore, PKC is also the most suitable choice
for another important application of cryptography, that is,
digital signature \cite{S96}.

The security of PKC lies on computational complexity assumptions,
which is reflected by the condition (C3). Equivalently, the
reliability of a PKC scheme is based on certain mathematically
difficult problems such as integer factorization, discrete
logarithm, etc. However, most of such problems are not difficult
in the context of quantum computation anymore \cite{S94,G96}. As a
result, most of PKC schemes will be broken by future quantum
computer. It is natural to ask, at that time, what is the
substitution for PKC to distribute a key? One possible way is to
exploit quantum mechanics, which is called quantum key
distribution (QKD) or quantum cryptography \cite{GRTZ}. QKD has a
unique property, that is, the potential eavesdropping would be
exposed by the users, and consequently it can achieve
unconditional security in theory. This security is assured by
fundamental principles in quantum mechanics instead of hardness of
computational problems.

In fact, QKD can only realize one application of PKC, i.e., key
distribution. But about digital signature, the other important
application, what can we do? Obviously we do not want to give up
the significant flexibility of PKC even in the era of quantum
computer. To this end the research is progressing along two
directions. One is to look for difficult problems under quantum
computation (especially the existing quantum algorithms
\cite{S94,G96}) and construct PKC based on them
\cite{OTU00,KKN05,Y03,K07}. In these schemes the key is still
composed of classical bits, and it follows that the flexibility of
PKC is retained. But the fact that their security lies on unproved
computational assumptions is unchanged. For simplicity, we call
this kind of cryptosystems the first class of quantum PKC (QPKC
class I). The other direction pursues PKC with perfect security by
adding more quantum elements in the schemes, which is just like
that of QKD \cite{N08,G05}. In these schemes the security is
assured by physical laws instead of unproved assumptions. However,
the keys generally contain qubits, which are, at least within
current techniques, more difficult to deal with, and then the
flexibility of PKC would be reduced to some extent. We call these
cryptosystems the second class of quantum PKC (QPKC class II). In
our opinion, both classes of QPKC are of significance for the
future applications. Class I is more practical, whereas class II
is more ideal and still needs more related researches. In this
paper we study the latter.

Recently, G. M. Nikolopoulos presented a novel QPKC scheme (GMN
scheme) based on single-qubit rotations \cite{N08}. In this scheme
the public key consists of polarization qubits. Each qubit is
generated by rotating a standard state $|0\rangle$ by a random
angle. All these angles (represented by bits) form the
corresponding private key. According to Holevo's theorem
\cite{NC00}, little information can be elicited by measuring these
qubits even when many copies of public key are served, which is
far from obtaining the exact value of the corresponding private
key. This basic idea is similar to the one proposed by Gottesman
\cite{G05}.

In this paper we will point out a potential security problem in
GMN scheme, and propose a new theoretical framework for QPKC based
on quantum encryption \cite{ZLG01,KBB02,BK03}. In our scheme two
qubits from a Bell state serve as the public key and the private
key respectively. Because both qubits are in the maximally mixed
state, we actually construct a quantum asymmetric cryptosystem
with symmetric keys, which seems unbelievable in conventional
cryptography. It is the quantum nature that renders this
interesting thing feasible. Furthermore, the security of this
scheme is guaranteed by physical laws in quantum mechanics. This
paper is arranged as follows. In Section II, we discuss the
security issue of previous idea of QPKC. Our new scheme is
presented in Section III and its security is analyzed in Section
IV. Finally some features of our scheme are discussed and
conclusions are drawn in Section V.

\section{A security issue in previous QPKC}
Let us briefly describe the preparation of the keys in GMN scheme
first. The user, say Bob, randomly chooses an integer $n$, and
then chooses $N$ integers $s_1,s_2,...,s_N$ from
$\mathbb{Z}_{2^n}$ independently, which compose an integer string
$\textbf{s}=(s_1,s_2,...,s_N)$. After that Bob generates $N$
single qubits in the states
$\{\hat{\mathcal{R}}^{(j)}(s_j\theta_n)|0\rangle\}$, where $1\leq
j\leq N$, $\theta_n=\pi/2^{n-1}$ and $\hat{\mathcal{R}}$ is the
rotation operation. It can be seen that these qubits are in
one-to-one correspondence with all the integers in $\textbf{s}$.
The private key is $d=\{n,\textbf{s}\}$, and the public key is
$e=\{N,|\Psi^{(PK)}_s(\theta_n)\rangle\}$, where
$|\Psi^{(PK)}_s(\theta_n)\rangle$ represents the state of the
sequence of all above $N$ qubits.

As analyzed in Ref. \cite{N08}, the entropy of the private key is
relatively high when $n\gg1$, and becomes higher with the
increasing of $n$. On the contrary, if an eavesdropper, say Eve,
wants to extract information about the private key by measuring
the public key (i.e. the qubits), she can only obtain limited
information. The obtainable information is bounded by Holevo
quantity \cite{NC00}, which totally depends on the number of the
qubits being measured. Therefore, it seems like that as long as
$n$ is large enough Bob can release many copies of his public key
without losing the confidentiality of his private key (see Eq.(3b)
in Ref.\cite{N08}).

From theoretical point of view, above conclusion is undoubtedly
right. However, when the security of the QPKC system is concerned,
the publication of multiple copies of public key would give Eve
the chance to attack. In fact, though knowing the private key
makes the eavesdropping very easy in QPKC, Eve's ultimate aim is
to obtain the encrypted message (i.e. the plaintext) instead of
the private key. Consequently, a straightforward strategy for Eve
arises, that is, trying to estimate the private key to certain
accuracy by measuring the public key and using the result to
obtain plaintext from the ciphertext.

Now we show what Eve can obtain by above strategy. To see the
particular accuracy to which Eve can estimate the private key, we
can use some results in the research of state estimation
\cite{MP95,DBE98,BBM02}. In GMN scheme, all the single-qubit
states lie on the $x$-$z$ plane of Bloch sphere. In this condition
by optimal collective measurements the obtainable fidelity between
the estimation result and the object state is \cite{DBE98,BBM02}
\begin{eqnarray}
F=\frac{1}{2}+\frac{1}{2^{M+1}}\sum_{i=0}^{M-1}\sqrt{\left(\begin{array}{c}M\\
i\end{array}\right)\left(\begin{array}{c}M\\i+1\end{array}\right)}\approx1-\frac{1}{4M}
\end{eqnarray}
where $M$ denotes the number of copies of the object state. That
is to say, if Eve has $M$ identical unknown states $|\psi\rangle$
on the $x$-$z$ plane, she can obtain a known state
$|\psi^{'}\rangle$ so that
\begin{eqnarray}
|\langle\psi|\psi^{'}\rangle|^2=F
\end{eqnarray}
It can be see that the guessed state $|\psi^{'}\rangle$ will be
very close to the object state $|\psi\rangle$ when $M$ is large.

Suppose Eve can get $K$ public keys in GMN scheme. Without loss of
generality, we take one state $|\psi_{s_j}\rangle$ as our example.
In this condition Eve has $K$ identical qubits in this state. Thus
she can obtain a guessed state $|\psi_{s_j}^{'}\rangle$ by optimal
collective measurements so that
$|\langle\psi_{s_j}|\psi_{s_j}^{'}\rangle|^2\approx1-1/(4K)$. Note
that here state $|\psi_{s_j}^{'}\rangle$ is known to Eve and it
means an approximate value of the integer $s_j$ in the private
key. As a result, Eve can construct a measurement basis
$B_{s_j}=\{|\psi_{s_j}^{'}\rangle,|\psi_{s_j}^{'\perp}\rangle\}$
and measure any single qubit in it ($|\psi_{s_j}^{'\perp}\rangle$
is the state orthogonal with $|\psi_{s_j}^{'}\rangle$). In the
following we will show that this basis brings Eve the chance to
extract information of the plaintext.

In the process of encryption the sender (say Alice) will get a
copy of Bob's public key, and use the qubit in state
$|\psi_{s_j}\rangle$ to encrypt the $j$th bit of her plaintext
$m_j$ ($m_j=$0 or 1). The corresponding ciphertext is the quantum
state $\hat{\mathcal{R}}^{j}(m_j\pi)|\psi_{s_j}\rangle$, which
implies that the plaintext 0 and 1 will be encrypted into the
ciphertext $|\psi_{s_j}\rangle$ and $|\psi_{s_j}^{\perp}\rangle$
respectively. Thus Eve can intercept the ciphertext sent by Alice
and measure it in the basis $B_{s_j}$, concluding the results
$|\psi_{s_j}^{'}\rangle$ and $|\psi_{s_j}^{'\perp}\rangle$
represent the plaintext 0 and 1 respectively. Since
$|\psi_{s_j}^{'}\rangle$ and $|\psi_{s_j}\rangle$ might be very
close on the Bloch sphere, Eve will obtain the correct plaintext
$m_j$ with a high probability, i.e. $P_c=F$. Equivalently, the
amount of the information Eve can obtained about $m_j$ equals
\begin{eqnarray}
I(A,E)&=&H(A)-H(A|E)\nonumber\\&=&1-2[F\log F+(1-F)\log(1-F)]
\end{eqnarray}
where A and E represent Alice and Eve respectively, and $I(A,E)$
denotes the mutual information between them.

Though the detection of eavesdropping is not involved in Ref.
\cite{N08}, we has to consider the disturbance brought by Eve's
intervention in view of its importance in quantum cryptography. In
above attack, to avoid being discovered by Alice and Bob, Eve can
resend her measurement result $|\psi_{s_j}^{'}\rangle$ or
$|\psi_{s_j}^{'\perp}\rangle$ to Bob after the measurement. In
this condition an error occurs with the probability
\begin{eqnarray}
P_e=2F(1-F)
\end{eqnarray}
Here ``error'' means the case where the bit sent by Alice is
different from the one received by Bob.

From Eqs.(3) and (4) it can be seen that Eve can obtain nearly all
the plaintext and, at the same time, introduce few errors when $K$
is large (please see Tab.1 and Fig.1 for details).

\begin{center}
\small Tab.1: The values of $I(A,E)$ and $P_e$ with different $K$.
\vspace{0.8mm}
\renewcommand\arraystretch{1.5}
\begin{tabular}{cccccc} \hline
& $K$=10 & $K$=20 & $K$=50 & $K$=100 & $K$=1000
\\\hline
$I(A,E)$ & 0.6627 & 0.8061 & 0.9092 & 0.9496 & 0.9933
\\
$P_e$\hspace{1mm} & \hspace{1mm}0.0488\hspace{1mm} &
\hspace{1mm}0.0247\hspace{1mm} & \hspace{1mm}0.0100\hspace{1mm} &
\hspace{1mm}0.0050\hspace{1mm} & \hspace{1mm}0.0005
\\\hline
\end{tabular}
\end{center}

\begin{figure}
\vspace{5mm}
\includegraphics[height=1.4in,width=2.4in]{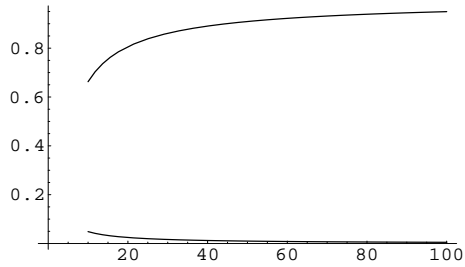}
\caption{Mutual information $I(A,E)$ and error probability $P_e$
as functions of the amount of the public key $K$. The horizontal
axis represents the values of $K$. The upper line and the lower
line indicate $I(A,E)$ and $P_e$, respectively.}
\end{figure}

Finally, in Ref. \cite{N08} and its very recent Erratum
\cite{N208}, it was pointed out that each bit in the plaintext
should be encrypted into several or more qubits so that this
scheme can stand against the SWAP-test attack. In this condition
our attack strategy may not be so effective either. However, as a
kind of special strategy to attack QPKC system, the
state-estimation attack seems more straightforward and practical
than the SWAP-test attack. The basic idea of our attack, i.e.,
estimating the state in public key by measurements and then trying
to decrypt the ciphertext instead of to recover the particular
private key, should be paid attention to in QPKC schemes including
Refs. \cite{N08,G05}. As a result, the state-estimation attack is
still of significance when the security of QPKC is concerned in
future research.

\section{QPKC based on quantum encryption}
From above discussion we can see that the model of key generation
in previous QPKC schemes may be vulnerable to the state-estimation
attack. More concretely, though Eve cannot obtain the exact
private key by measuring multiple copies of the public key, she
can still get an approximate private key and then use it to elicit
information about the plaintext. Therefore, it would be desirable
to find a new way to generate keys in QPKC. Here we will give a
scheme using the qubits from Bell state as keys, in which, as in
almost all existing protocols of quantum cryptography, the process
of eavesdropping detection is introduced and the security is
guaranteed by it.

Before the description of our QPKC scheme, it is necessary to
introduce several basic assumptions about this system. That is,
(A1) there is a believable center (Trent) in the QPKC system; (A2)
Trent can authenticate every user's identity in the communications
between them, which can be realized by quantum authentication
protocols \cite{QA4}; (A3) the information transmitted in the
classical channels can be eavesdropped, but cannot be modified.
These assumptions are reasonable and generally accepted in PKC
(e.g. A1 and A2) and quantum cryptography (e.g. A3).

This scheme consists of the following four stages.

\emph{Stage 1: Key generation.} Trent generates a pair of keys,
i.e., the public key $e$ and the private key $d$, for each user.
Without loss of generality, consider Bob as our example. The
particular process is as follows.

1. Trent prepares a sequence of qubit pairs
$S_1=\{(p_1,q_1),(p_2,q_2),...,(p_n,q_n)\}$. Each pair is in the
Bell state
\begin{eqnarray}
|\Phi^+\rangle=\frac{1}{\sqrt{2}}(|00\rangle+|11\rangle)
\end{eqnarray}
Two qubit sequences $S_p=\{p_1,p_2,...,p_n\}$ and
$S_q=\{q_1,q_2,...,q_n\}$ will be used as Bob's public key and
private key, respectively.

2. To securely transmit $S_q$ to Bob, Trent also generates a
certain quantity of decoy states $S_d=\{d_1,d_2,...,d_k\}$, where
every qubit is randomly in one of the states $\{|0\rangle,
|1\rangle, |+\rangle=\frac{1}{\sqrt{2}}(|0\rangle+|1\rangle),
|-\rangle=\frac{1}{\sqrt{2}}(|0\rangle-|1\rangle)\}$. Please note
that here the meaning of \emph{decoy state} is somewhat different
from that, as widely studied now \cite{Decoy}, used in the way to
resolve the problem of Photon-Number-Splitting (PNS) attack in a
practical QKD implementation. However, the tasks of both kinds of
decoy states are the same, that is, helping users discover
potential attacks. We will discuss the role of above decoy states
in detail in Section IV.

3. Trent inserts each qubit in $S_d$ into a random position of the
sequence $S_q$, obtaining a new qubit sequence $S_{qd}$. Then
Trent sends $S_{qd}$ to Bob via a quantum channel.

4. After Bob received all these qubits, Trent tells Bob the
position and the basis (i.e. $B_z=\{|0\rangle,|1\rangle\}$ or
$B_x=\{|+\rangle,|-\rangle\}$) of each decoy state.

5. Bob measures all decoy states in their corresponding bases, and
then announces the measurement results to Trent. By comparing
these results with the initial states of these qubits, Trent can
judge whether the transmitted sequence is disturbed.

6. If no eavesdropping occurs, Bob obtains his private key $d$,
i.e. the sequence $S_q$. At the same time, Trent stores Bob's
public key $e$, i.e. $S_p$, for future usage. Otherwise the
communication may be insecure and will abort.

In the following stages we can see that the keys might be not
enough for encrypting a long message, or be consumed gradually.
But whenever it is not enough to be used, Trent can generate new
Bell-state pairs to refuel the keys.

\emph{Stage 2: Encryption.} Suppose a user, say Alice, wants to
send an $r$-bit message $m=\{m_1,m_2,...,m_r\}$ to Bob, where
$m_i=0$ or 1, and $r\leq n$. Then Alice can encrypt it according
to the following steps.

1. Alice requests Trent to send her $r$ qubits of Bob's public
key.

2. Trent sends the first $r$ qubits of the sequence $S_p$ to
Alice. Here we use $S_p^r$ to denote this part of sequence, i.e.
$S_p^r=\{p_1,p_2,...,p_r\}$. Similar to that in Stage 1, Trent
also utilizes decoy states so that these qubits are securely
transmitted to Alice.

3. Alice generates an $r$-qubit sequence $L=\{l_1,l_2,...,l_r\}$
with states $\{|m_1\rangle,|m_2\rangle,...,|m_r\rangle\}$
respectively, which corresponds to her message to be encrypted.

4. Alice encrypts her message $L$ with the public key $S_p^r$.
More concretely, Alice uses one qubit in $S_p^r$ to encrypt her
corresponding message qubit via a CNOT operation. For example, to
encrypt $|l_i\rangle$, Alice performs a CNOT gate $C_{p_il_i}$
(the first subscript $p_i$ denotes the controller and the second
$l_i$ represents the target) on qubits $p_i$ and $l_i$, that is
\begin{eqnarray}
C_{p_il_i}|\Phi^+\rangle_{p_iq_i}|m_i\rangle_{l_i}=\frac{1}{\sqrt{2}}(|00m_i\rangle+|11\overline{m}_i\rangle)_{p_iq_il_i}
\end{eqnarray}
where $\overline{m}_i=1-m_i$.

5. After the encryption of all her message qubits, Alice sends the
sequence $L$ (the ciphertext) to Bob through a quantum channel.

\emph{Stage 3: Decryption.} After Bob received all these qubits,
he can execute the following steps to recover the message $m$.

1. For each qubit in the ciphertext $L$, Bob performs a CNOT
operation $C_{q_il_i}$ to decrypt it. Then the state changes into
\begin{eqnarray}
C_{q_il_i}\frac{1}{\sqrt{2}}(|00m_i\rangle+|11\overline{m}_i\rangle)_{p_iq_il_i}=|\Phi^+\rangle_{p_iq_i}|m_i\rangle_{l_i}
\end{eqnarray}

2. Bob measures each qubit in $L$ in basis $B_z$. From Eq.(7) we
can see that the measurement results exactly compose the message
$m$. Thus the message sent by Alice is recovered and the
decryption is finished.

\emph{Stage 4: Key recycling.} There is a good property in the
above communication, that is, the states of Bob's keys are still
unchanged after the processes of encryption and decryption.
Therefore, the keys can be recycled according to the following
steps.

1. Alice sends Bob's public key, i.e. the qubit sequence $S_p^r$
to Trent.

2. To ensure the security of these recycled key qubits, Trent
randomly selects a certain number of them from $S_p^r$ as the test
qubits, and measures each of them in $B_z$ or $B_x$ at random.

3. Trent tells Bob the position and the measurement basis of each
test qubit.

4. Bob measures his corresponding qubits in the same bases and
announces his results. Because every two corresponding qubits in
two keys should be in Bell state $|\Phi^+\rangle$, the measurement
results would exhibit deterministic correlations. For example,
they are equal in the measurement in both bases $B_z$ and $B_x$.

5. By comparing their measurement results Trent can judge whether
these qubits are attacked. If they are not, Trent and Bob store
the remaining qubits to refuel the public key and the private key.
Otherwise the recycled key qubits would be discarded.

Now we have described the QPKC scheme based on quantum encryption.
It can be seen that both the qubits in public key and the ones in
private key come from Bell state $|\Phi^+\rangle$, and are in the
same state (i.e. the maximally mixed state
$\rho=1/2(|0\rangle\langle0|+|1\rangle\langle1|)$). Therefore, an
interesting event happens. That is, this QPKC scheme essentially
use a pair of \emph{symmetric} keys. In fact the basic idea of
this scheme is similar to that of quantum Vernam cipher
\cite{L02}. In conventional cryptography, as we know, the Vernam
cipher (i.e. one-time pad) \cite{V26} can never be used in PKC
because its decryption key and encryption key are equal, and they
can be copied at will. But in the quantum context things become
totally different. That is, one cannot obtain the decryption key
(i.e. private key) by replicating a copy of the encryption key
(i.e. public key) even though they are in the same state, which is
guaranteed by quantum no-cloning theorem \cite{WZ82}.

Finally, about this QPKC scheme, there are some issues to be
clarified.

1. In fact the public key obtained by Alice is a sub-sequence of
$S_p$. After Alice received these qubits, it is necessary for
Trent to tell Bob which sub-sequence of $S_p$ was sent to Alice so
that Bob can use his corresponding qubits to decrypt Alice's
ciphertext. By this way Bob can correctly decrypt every ciphertext
even though there are multiple ciphertexts received simultaneously
from different senders.

2. In above description Alice and Bob do not detect the potential
eavesdropping to the ciphertext, which may happen when it was
transmitted in the channel. As we will show in Section IV, Eve
cannot obtain the message from the ciphertext. But she can still
do a denial-of-service (DoS) attack to disturb the communication
\cite{C03,GGWZ08}. To enable Alice and Bob to discover this kind
of attack, the method of message authentication can be introduce
to this scheme. For example, when Alice wants to send message $m$
to Bob, she computes the message digest $H(m)$ via a public Hash
function (e.g. MD5, SHA-1, et al.) \cite{S96} first, and then
sends both $m$ and $H(m)$ to Bob by above QPKC system. Thus after
Bob received the corresponding two parts $m'$ and $H'(m)$ he can
detect eavesdropping by verifying whether $H'(m)$ is the message
digest of $m'$. By this means Alice and Bob can discover the
eavesdropping to the ciphertext.

3. Till now we have not consider the noise in quantum channels. As
we know, quantum state will change because of the unavoidable
decoherence in a noisy channel. In this condition, the
technologies of entanglement purification \cite{BBP96,PGU03} and
quantum privacy amplification \cite{DEJ96} can be introduced in
this scheme to improve the quality of the Bell state of these EPR
pairs (i.e. the keys) after the transmission of one of the keys.
Therefore, our QPKC scheme can be used even in a noisy
circumstances.

\section{Security analysis}
In a QPKC system the aim of Eve is to obtain Bob's private key,
which can be used to decrypt the ciphertext, or alternatively,
obtain the plaintext without the private key. Therefore, it must
be ensured that the above two events cannot occur in a secure QPKC
system. The following discussions will be based on this fact.

In above QPKC scheme some familiar and reliable manners are
utilized to guarantee its security. For example, BB84-type qubits
\cite{BB84} are used as the decoy states to protect the
transmitted sequence, and conjugate-bases measurements to identify
the state of recycled key qubits. Note that in our scheme every
public-key qubit is only used to encrypt one message bit (or
qubit), so there is no correlation between different ciphertexts.
As a result, we have no need to consider the conventional attack
strategies such as chosen-plaintext attack and chosen-ciphertext
one. In the following we will briefly discuss the security with
respect to different stages of this scheme.

\emph{Key generation.} In this stage Trent prepares EPR pairs in
$|\Phi^+\rangle$ and sends one qubit in each pair (i.e. the
sequence $S_q$) to Bob as his private key. Because Trent is
believable we only need to consider the attack from an outside
eavesdropper (Eve). In this process Eve has the chance to obtain
Bob's private key, with which she can decrypt any ciphertext sent
to Bob. However, Eve's goal will not be achieved because of the
usage of decoy states. The reasons are as follows.

First, quantum no-cloning theorem \cite{WZ82} ensures that Eve
cannot replicate the qubits in the private key. For simplicity,
consider one EPR pair $(p_i,q_i)$, where $p_i$ is a qubit in
sequence $S_p$ (public key) and $q_i$ is the one in $S_p$ (private
key). Obviously one cannot generate a new qubit $q^{'}_i$, a copy
of $q_i$, when $q_i$ is transmitted in the channel so that both
$(p_i,q^{'}_i)$ and $(p_i,q_i)$ are in Bell state
$|\Phi^+\rangle$. This is guaranteed by fundamental laws in
quantum mechanics. This point is very different from that in
conventional PKC systems, in which the private key can never be
transmitted in the public channel because it is in the form of
bits and can be easily copied.

Second, since both the decoy qubits and the private-key ones are
in the same state, i.e. the maximally mixed state
$\rho=1/2(|0\rangle\langle0|+|1\rangle\langle1|)$, these two kinds
of qubits cannot be distinguished. That is to say, any attack
operation which is expected to be performed on the private-key
qubits will be also inevitably executed on the decoy ones. As a
result, the attack would leave a trace on the decoy states and
then be discovered by legal users. For example, Eve may want to
entangle her ancilla into the Bell state by a collective operation
on it and qubit $q_i$, and subsequently use the ancilla to decrypt
the ciphertext which was encrypted by $q_i$. More concretely, Eve
prepares an ancilla $|0\rangle_a$, and performs a CNOT operation
$C_{q_ia}$ when $q_i$ is transmitted in the channel. That is,
\begin{eqnarray}
C_{q_ia}\frac{1}{\sqrt{2}}(|00\rangle+|11\rangle)_{p_iq_i}|0\rangle_a=\frac{1}{\sqrt{2}}(|000\rangle+|111\rangle)_{p_iq_ia}.
\end{eqnarray}
And then resends $q_i$ to Bob. When Alice uses $p_i$, the
corresponding public-key qubit, to encrypt a message bit $m_i$,
the state of the whole system changes into
\begin{eqnarray}
C_{p_il_i}\frac{1}{\sqrt{2}}(|000\rangle&&\hspace{-3.5mm}+|111\rangle)_{p_iq_ia}|m_i\rangle_{l_i}\nonumber\\&&\hspace{-3.5mm}=\frac{1}{\sqrt{2}}(|000m_i\rangle+|111\overline{m}_i\rangle)_{p_iq_ial_i}.
\end{eqnarray}
In this condition Eve can correctly obtain $|m_i\rangle$ if she
intercepts $l_i$, the ciphertext qubit, when it is transmitted to
Bob and performs the following operation
\begin{eqnarray}
C_{al_i}\frac{1}{\sqrt{2}}(|000m_i\rangle&&\hspace{-3.5mm}+|111\overline{m}_i\rangle)_{p_iq_ial_i}\nonumber\\&&\hspace{-3.5mm}=\frac{1}{\sqrt{2}}(|000\rangle+|111\rangle)_{p_iq_ia}|m_i\rangle_{l_i},
\end{eqnarray}
which means Eve gets the plaintext $m_i$.

However, the above attack will bring disturbance to the decoy
states. For example, consider decoy state $|+\rangle$. When Alice
intercepts it and performs her first CNOT operation on it and her
ancilla $|0\rangle$, they will come into Bell state
$|\Phi^+\rangle$, which results in a totally random result when
Bob measures the decoy state to detect eavesdropping.

Therefore, the above attack will be inevitably discovered by Bob
and Trent. In fact, BB84-type particles can reliably guarantee the
security of a quantum sequence, which has been reflected by the
proved security of BB84 protocol \cite{BB84,LC99,SP00}.
Equivalently, any effective attack will be disclosed by the
detection via those particles.

\emph{Encryption}. As introduced in Section III, we use symmetric
keys in our QPKC scheme. That is, the public key and the private
one are in the same state. Therefore, anyone who has the public
key can also decrypt the ciphertext encrypted by this key. In this
stage Eve has the chance to touch the public key when it is
transmitted from Trent to Alice. However, similar to that in Stage
1, Eve can never replicate those qubits and the decoy states
ensure the security of the public key. Consequently, any effective
attack on the public key will be discovered by legal users.

Now let us observe what Eve can obtain from the ciphertext when it
is transmitted from Alice to Bob. From above analysis, it can be
seen that Eve cannot elicit any helpful information from the
transmitted key qubits, including both the public key and the
private key, if she does not want to bring disturbance to the
decoy states. In this condition Eve can obtain nothing about the
plaintext from the ciphertext because all ciphertext qubits are in
the same state $\rho=1/2(|0\rangle\langle0|+|1\rangle\langle1|)$
in spite of the value (0 or 1) of corresponding message bit.

The classical Hash function is used in this stage. We should
emphasize that, though Hash functions are not perfectly secure
(e.g., collisions might be found by some advanced algorithms
\cite{WXY04,WXY05}), it does not decrease the security of the
whole QPKC system. In this stage, as shown above, Eve cannot
obtain the plaintext at all. The usage of Hash function is just to
protect the scheme against DoS attack. In fact it plays the role
like message authentication code (MAC). As a result, general Hash
functions such as MD5, SHA-1, et al. can disclose a potential DoS
attack.

\emph{Decryption}. In this stage Eve has no chance to attack
because no qubits are transmitted in the channel. After Bob
obtained the plaintext, he can judge whether DoS attack occurred
with the help of Hash function.

\emph{Key recycling.} In this stage Alice sends the public key
back to Trent. This situation, as far as Eve is concerned, is
similar to that in the beginning of Stage II. But here we should
also consider the attack from Alice. Because the recycled
public-key qubits will be reused in later applications where
another one (say Charlie) sends his message to Bob, Alice can do
something for future illegal decryption when these qubits are
still in her hand. For example, Alice can entangle her ancilla
into each Bell states and use it to decrypt the ciphertext sent by
Charlie later (similar to Eve's strategy in Stage 1 and the ones
in Refs. \cite{GGW1,GGW2}).

Taking above threat into account, we have to ensure that the
states of the public-key qubits Alice sent back are unchanged
(that is, each qubit is still in Bell state $|\Phi^+\rangle$ with
its corresponding particle in Bob's hand). In our scheme we use
the manner of conjugate-bases measurements to detect
eavesdropping, which can resist attacks from both Eve and Alice.
This manner has been widely used in quantum cryptography and its
reliability has been proved \cite{BBM92,WZY02}. Here we will not
repeat the analysis any more.

Finally, it is well known that, in a practical QKD system, Eve may
attack only a little part of the transmitted particles so that the
introduced disturbance will be covered up by channel noises. In
this case Eve can elicit a small amount of information about the
key. And at the same time, legal users cannot ascertain whether
there is an eavesdropper in the channel because the error rate
introduced by Eve is small enough. At that time, the users can
perform privacy amplification \cite{BBR88,BBC95} on the raw key
and then obtain a final key with unconditional security. In our
QPKC scheme, similar problem also exists. Eve may attack only a
little part of the key qubits and then obtain some information
about the plaintext. In this condition we introduce entanglement
purification \cite{BBP96,PGU03} and quantum privacy amplification
\cite{DEJ96} in our scheme, which makes it possible to achieve
unconditional security in theory.

\section{Discussions and conclusions}
Compared with the previous QPKC system (GMN scheme) \cite{N08},
our scheme has the following features.

1. The roles of public key and private key are equal. When Rivest,
Shamir, and Adleman presented the famous RSA scheme \cite{RSA78},
as described in Section I, they pointed out four basic conditions
which a PKC system generally satisfies. Among them the last
condition (C4) requires that the users can also use private key to
encrypt a message and use public key to decrypt it correctly. This
requirement opens the door for an important application of PKC,
i.e. digital signature. But this aim is not achieved in GMN
scheme. The problem is resolved in our scheme because both public
key and private key are quantum one and in the same state.
Therefore, \emph{this feature makes it possible to construct a
quantum signature protocol based on our scheme}. Of course to
design such a protocol is a complex work \cite{GC01,Zeng02} and it
is beyond the scope of this paper.

2. The manner to verify the identity of public key is presented in
our scheme. In both schemes public key is quantum one and its
identity should be authenticated when the message sender received
it from Trent (or a key-distribution center, i.e. KDC, called in
Ref. \cite{N08}). This is a crucial point for the security of
whole QPKC system. However, authentication is still an open
question in GMN scheme because of the complexity of the public-key
states. In our scheme this problem is resolved from two aspects.
On the one hand, the decoy-states detection is utilized to protect
the public-key qubits from being attacked by Eve. On the other
hand, because the key qubits are from the same Bell state
$|\Phi^+\rangle$, entanglement purification and quantum privacy
amplification can be easily performed on them in the sense that
they are existing technologies for Bell states
\cite{BBP96,PGU03,DEJ96}. Through these manners high-fidelity Bell
state can be finally obtained even under a noisy channel, or
equivalently, the state of public key can be authenticated. On the
contrary, the public-key states are different from each other and
even unknown for the message sender in GMN scheme, which makes it
very hard to perform quantum privacy amplification on them.

3. The state-estimation attack is invalid for our scheme. In GMN
scheme, as discussed in Section II, Eve can estimate the state of
public key by measuring multiple copies of them, and then obtain
much information about the plaintext. However, in our scheme any
two qubits from different public key belong to different EPR pairs
and there are no correlations between them. Even though the same
qubit is reused in subsequential encryption, it is independent
with itself in previous usage because its state is identified in
the process of recycling. As a result, Eve cannot get more useful
information from multiple public key than that from one. In fact,
as pointed out in Section IV, no one can obtain a copy of private
key (or qubits with which Eve can correctly decrypt a certain
ciphertext) from the public key without introducing disturbance.
This is guaranteed by fundamental laws in quantum mechanics.

4. The keys can be reused and refuelled whenever it is needed.

We have to confess that, apart from above features, there is also
a disadvantage of our scheme. That is, private key consists of
qubits in stead of bits as in GMN scheme, which presents a burden
to the user to store them. However this is not a fatal problem
because quantum storage seems necessary in a QPKC system. For
example, many copies of public key must be stored by Trent or KDC
for a long time.

One may argue that our scheme does not look like a practical PKC
system (e.g. any familiar conventional PKC such as the famous RSA
scheme) for the following two reasons: 1. Some QKD-like strategies
for eavesdropping detection are used to guarantee the security; 2.
It uses symmetric keys. We emphasize that \emph{all these facts
have their roots in the quantum nature of QPKC}. Now let us give
further interpretations about above two questions.

1. As we know, the quantum-mechanical nature of qubits renders
eavesdropping detectable, which is the root of the unconditional
security of quantum cryptography. To obtain this advantage in a
quantum protocol, an eavesdropping-detection process is absolutely
necessary. It is also the fact in QPKC. For example, in a QPKC
system the public key, generally composed by qubits \cite{FN1},
must be authenticated after the transmission in a public channel.
Otherwise Eve may correctly decrypt the corresponding ciphertext
by a prior attack on this key (e.g. replacing it with her own
qubits or entangling ancillas into it). Therefore, we have to
introduce some QKD-like strategies to protect the security of the
public key, which exactly reflects the essential characteristic of
quantum cryptograph. On the contrary, the classical public key in
RSA scheme can be easily authenticated by conventional
technologies such as digital signature \cite{S96}. Note that there
is no such strategies in GMN scheme because the content of
public-key authentication is not contained in Ref. \cite{N08}.

2. By choosing Bell-state qubits as the keys we initially intended
to avoid the state-estimation attack as in GMN scheme. In fact
Bell states have a special feature which is suitable for QPKC.
That is, these states can be authenticated by existing
technologies (especially entanglement purification and quantum
privacy amplification), which is an important issue in QPKC but
still not resolved in the previous scheme. We know that people can
never use equal keys in a conventional PKC system because in this
condition anyone can get the private key just by replicating a
copy of the public key, and then decrypt all corresponding
ciphertexts. Thus we really need to design two different keys so
that Eve cannot obtain the private key from the public one.
However, in the quantum circumstance, things go very differently.
On the one hand, quantum no-cloning theorem does not allow the
replication of qubits any more. On the other hand, the
authentication of public key is necessary in QPKC and, at the same
time, whenever the authentication is successful it generally
ensures that Eve cannot read any information from public key.
\emph{In this condition, therefore, we have no need to design two
different keys any more}. That is, equal keys are competent for
QPKC. In fact we have shown that it is feasible to use symmetric
keys in QPKC system, which touches on the very nature of the
quantum state.

In conclusion, as a subsequent study of Ref. \cite{N08}, we gave a
new elementary idea for QPKC and constructed a whole theoretical
framework of a QPKC system. It was shown that symmetric keys could
be used in QPKC, which is quite different from that in
conventional PKC. The security and features of this scheme were
discussed. In addition, a possible attack to GMN scheme \cite{N08}
was demonstrated. Combining the unconditional security of QKD and
the significant flexibility of PKC, QPKC has been an expected goal
of the scholars in the field of quantum cryptography for a long
time. But to design a practical QPKC scheme, or alternatively, to
demonstrate its feasibility, is still a difficult work. This study
can be seen as a step towards this direction.

\section*{ACKNOWLEDGMENTS} This work is supported by the National High Technology
Research and Development Program of China, Grant No. 2006AA01Z419;
the Major Research Plan of the National Natural Science Foundation
of China, Grant No. 90604023; the National Laboratory for Modern
Communications Science Foundation of China, Grant No.
9140C1101010601; the Natural Science Foundation of Beijing, Grant
No. 4072020; and the ISN Open foundation.


\begin{thebibliography}{ss3}
\bibitem{FN1} In section I we divided QPKC into two classes. We believe that the public key in QPKC class II must
be quantum one. Otherwise a classical public key has not the
ability to detect eavesdropping, and then cannot obtain
unconditional security in theory, which is the characteristic of
QPKC class I.
\end{thebibliography}
\end{document}